\documentclass[11pt, a4paper]{article}
\usepackage{graphicx}
\usepackage{amssymb}
\usepackage{color}
\usepackage{setspace}
\usepackage[margin=1.0in]{geometry}

\begin{document}

\begin{figure*}[tb]
	\centering
	\includegraphics[width=1.00\textwidth]{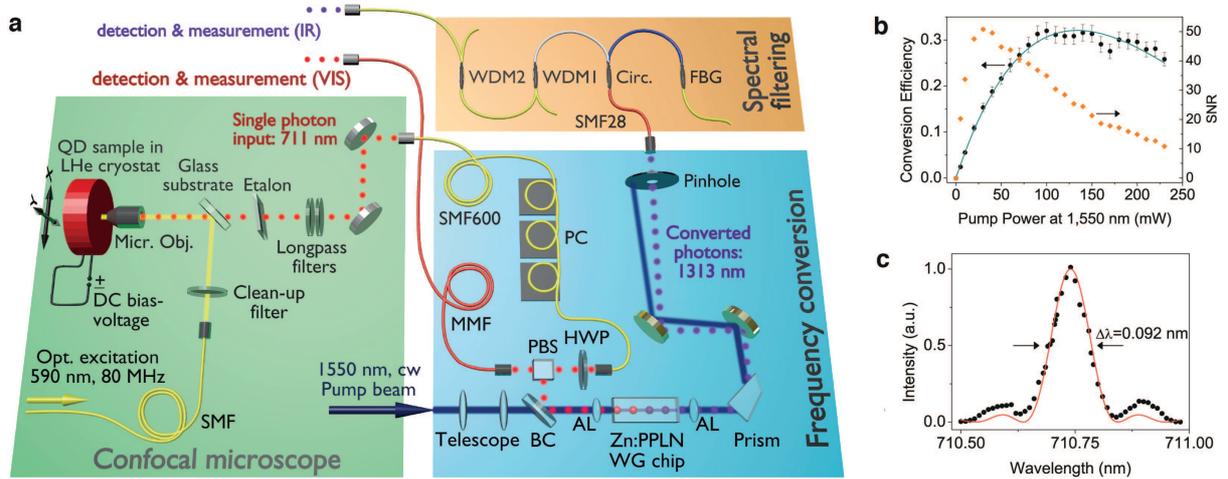}
	\caption{\textbf{Frequency downconversion of nonclassical light emitted by a single QD.} (a)~Experimental setup. The confocal microscope consists of a liquid helium (LHe) continuous flow cryostat containing the QD sample at a temperature of 12\,K. A bias-voltage of 3.20-3.47\,V is applied to the quantum dots which are additionally optically excited using an average power of 160~nW at 590~nm from a pulsed fs optical parametric oscillator (OPO) with 80~MHz repetition rate \cite{Ruebel2008}. Optical excitation and collection of photoluminescence (PL) emitted by a QD are performed using a $100\times$ microscope objective with numerical aperture NA = 0.8. Under the excitation conditions given above we collect a rate of $\sim \!\!\! 188,400$ photons/s into a single-mode fibre (SMF600) guiding them to the frequency conversion setup. A silica etalon can be inserted optionally for narrow-band filtering of the PL. Two longpass filters prevent residual excitation light from entering the fibre. For frequency down-conversion the visible photons are coupled into the Zn:PPLN ridge waveguide together with a strong pump beam at 1,550~nm provided by a continuous-wave OPO (see methods for details). The converted photons are spatially separated from the strong pump light and from residual visible photons by a prism and a pinhole and are coupled into a standard telecom fibre (SMF28). To suppress residual pump light and noise background around the target wavelength $\lambda_{\mathrm{out}}$ we additionally use a spectral filtering setup composed of a fibre-optic circulator, a fibre Bragg grating (FBG) centred at 1,312.714~nm ($-1.0$\,dB reflection bandwidth: 0.755\,nm) and two 1,310~nm/1,550~nm wavelength division multiplexers (WDM1 and WDM2). (b) Total efficiency and SNR of the frequency conversion setup calculated from measured data using single photon input from a QD. (c) Spectral acceptance bandwidth of the DFG process measured for 2\,mW input power around 710.74~nm from a tunable cw Ti:Sapphire laser (red line: $\mathrm{sinc}^2$-fit). For further details see Supplementary Information.}
	\label{fig:Setup}
\end{figure*}

\begin{figure}[H]
	\centering
	\includegraphics[width=0.5\textwidth]{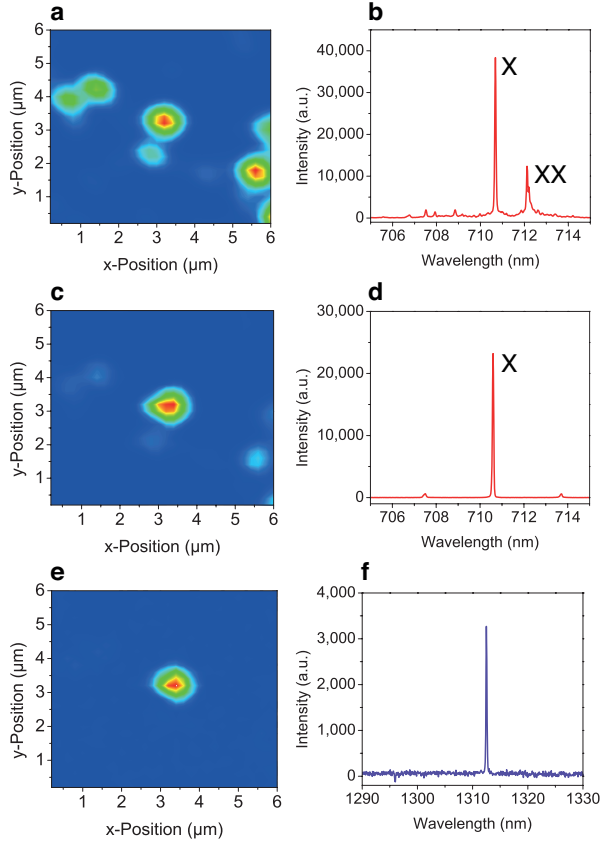}
	\caption{\textbf{Photoluminescence maps with corresponding spectra.} To analyse the spectrum of the visible QD emission we employ a grating spectrometer with a CCD detector. The spectrum of the 1,313~nm-signal can be recorded using another grating spectrometer with an InGaAs linear array detector. In both devices we select a grating with 1,800 lines/mm. (a), (c), (e) Same $6 \times 6\,\mathrm{\mu}$m detail of the QD sample. Maps (a) and (c) were recorded by scanning the x-y-position of the sample and directly detecting the visible PL from the quantum dot using a Si-APD without (a) and with (c) etalon filtering. Map (e) was obtained by scanning the same sample area but detecting the down-converted light at 1,313~nm with a SSPD. (b), (d), (f) Spectra that were measured by setting the x-y-position of the sample to the point corresponding to the maximum intensity of the central quantum dot. (b), (d) Visible PL spectrum without and with etalon filtering, respectively. The two prominent lines in (b), separated by 1.45\,nm (3.54\,meV), are attributed to exciton (X) and biexciton (XX) transitions. (f) Converted spectrum illustrating the effect of spectral filtering by a combination of DFG acceptance curve and FBG.}
	\label{fig:FluorescenceMaps}
\end{figure}

\begin{figure}[H]
\centering
\includegraphics[width=0.4\textwidth]{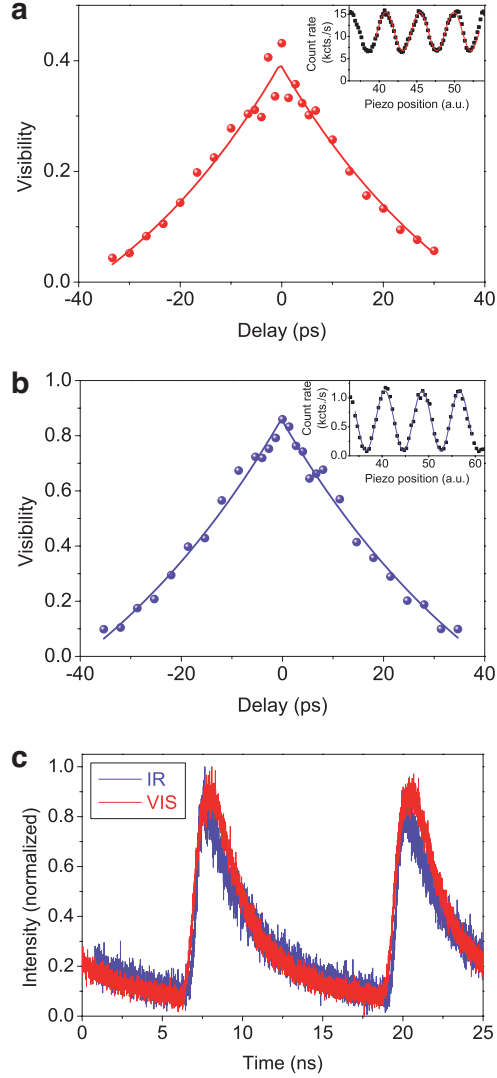}
\caption{\textbf{Single-photon temporal coherence and lifetime before and after frequency down-conversion.} (a) Visibility of first-order interference fringes vs. time delay for PL from the quantum dot at 711\,nm deduced from measurements of the $g^{(1)}(\tau)$ correlation function for different delay times $\tau$. The inset shows an example of interference fringes at fixed delay time. (b) Same measurement as (a) but for converted light at 1,313\,nm. (c) Lifetime of single photons measured via time-correlated single photon counting for original visible light and converted telecom light.}
\label{fig:g1measurement}
\end{figure}

\begin{figure}[H]
\centering
\includegraphics[width=0.4\textwidth]{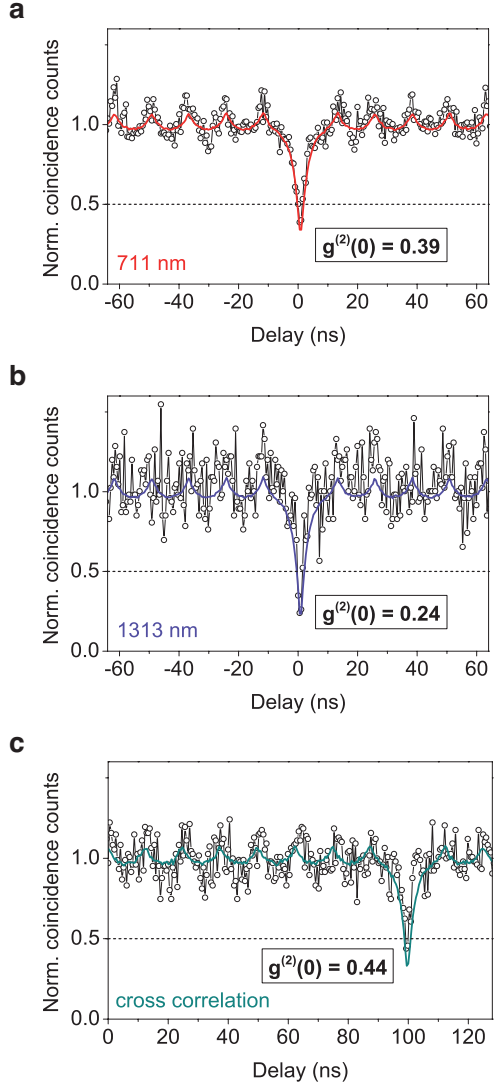}
\caption{\textbf{Conservation of photon antibunching under frequency downconversion: simulation and measurement.} (a) $g^{(2)}$ correlation function of the single photons at 711\,nm emitted by the quantum dot measured with a HBT setup for visible light. (b) $g^{(2)}$ correlation function of the converted light at 1,313~nm measured with a HBT setup for telecom light. (c) $g^{(2)}$ cross-correlation function measured with a hybrid HBT setup, i.e., one half of the visible photons is sent directly to a visible photon detector, the other half undergoes frequency down-conversion and is then detected with an IR photon detector. Detection events at the two individual detectors are anti-correlated. The fact that the dip occurs at $\tau^{\prime} \approx 100$~ns perfectly matches the difference of the path lengths the photons of different colour have to travel to the respective detectors. Black circles in Fig. (a)-(c) represent measured data while the red, violett and green curve, respectively, are obtained from Monte Carlo simulations (see Supplementary Information). All experimental and simulated data have been independently normalized with respect to the correlation function of a perfect cw Poissonian light source of equal average intensity, i.e., $g^{(2)}(\tau)=G^{(2)}(\tau)/(N_1 N_2 t_{\mathrm{bin}} t_{\mathrm{int}})$, where $G^{(2)}(\tau)$ represents the measured coincidence counts, $N_1$ and $N_2$ are the time-averaged count rates on detector 1 and 2, respectively, $t_{\mathrm{bin}}$ is the time bin width and $t_{\mathrm{int}}$ is the integration time \cite{Michler2009}. To ensure comparability, all measurements and simulations were performed with a time binning of $t_{\mathrm{bin}}=512$\,ps. This rather large bin size generally leads to an overestimation of the $g^{(2)}(0)$-values, i.e., for perfect timing resolution the measured data would correspond to $g^{(2)}_{\mathrm{vis}}(0) = 0.23$ ($g^{(2)}_{\mathrm{IR}}(0) = 0.15$) for the input (converted) photons, respectively. No background has been subtracted from the experimental data in all plots. See Methods for details.}
\label{fig:g2measurement}
\end{figure}

\begin{center}
\textbf{\Large{Visible-to-telecom quantum frequency conversion of light from a single quantum emitter}}\\
\vspace{10pt}
Sebastian Zaske$^1$, Andreas Lenhard$^1$, Christian A. Ke\ss{}ler$^2$, Jan Kettler$^2$,\\ Christian~Hepp$^1$, Carsten Arend$^1$, Roland Albrecht$^1$, Wolfgang-Michael Schulz$^2$, Michael Jetter$^2$, Peter Michler$^2$ and Christoph~Becher$^{1,}$\footnote{Corresponding author: christoph.becher@physik.uni-saarland.de, Tel: +49~(0)681 302 2466, Fax: +49~(0)681~302~4676}
\end{center}
\textit{1. Fachrichtung 7.2 (Experimentalphysik), Universit\"at des Saarlandes, Campus E2.6, 66123 Saarbr\"ucken, Germany\\
2. Institut f\"ur Halbleiteroptik und Funktionelle Grenzfl\"achen and Research Center SCoPE,
Universit\"at Stuttgart, 70569 Stuttgart, Germany}
\vspace{22pt}
\\
\textbf{Quantum frequency conversion (QFC), a nonlinear optical process in which the frequency of a quantum light field is altered while conserving its non-classical correlations, was first demonstrated 20 years ago \cite{Huang1992}. Meanwhile, it is considered an essential tool for the implementation of quantum repeaters \cite{Briegel1998, Sangouard2011, Ou2008, Takesue2010, Curtz2011} since it allows for interfacing quantum memories with telecom-wavelength photons as quantum information carriers. Here we demonstrate efficient ($>30$\%) QFC of visible single photons (711\,nm) emitted by a quantum dot (QD) to a telecom wavelength (1,313\,nm). Analysis of the first and second-order coherence before and after wavelength conversion clearly proves that important properties, such as the coherence time and photon antibunching, are fully conserved during the frequency translation process. Our findings underline the great potential of single photon sources on demand in combination with QFC as a promising technique for quantum repeater schemes.}\\
\noindent The majority of QFC experiments \cite{Albota2004, Roussev2004, Tanzilli2005, Rakher2010, Pelc2011} that have been performed so far focused on quantum frequency up-conversion to efficiently translate telecom light to visible wavelengths. In a pioneering work Tanzilli \textit{et al.} \cite{Tanzilli2005} demonstrated the conservation of photon-entanglement under frequency up-conversion. Furthermore, it was experimentally proven that photon antibunching of light from a single quantum emitter is not destroyed by this nonlinear process \cite{Rakher2010}. In contrast to up-conversion interest in the reverse process, quantum frequency down-conversion, began to grow only in recent years \cite{Ou2008, Takesue2010, Curtz2011, McCutcheon2009, Pelc2010, Zaske2011}. Approaches in this field are driven by the insight that long-haul quantum networks will require telecom photons as flying qubits while the most efficient quantum memories to date operate at wavelengths around 800~nm. In this respect two seminal experiments have been reported recently demonstrating down-conversion of weak non-classical light fields \cite{Radnaev2010} and entangled photons \cite{Ikuta2011}. Considering proposed quantum repeater schemes, such as the Duan-Lukin-Cirac-Zoller (DLCZ) protocol \cite{DLCZ} and derivates thereof \cite{SimonIrvine2003, Simon2007, Sangouard2007}, a significant increase of the entanglement distribution rate could be achieved by using on-demand single photon sources instead of photon pair sources \cite{Sangouard2007}. In this context, bridging the gap between the system wavelengths of quantum memories/processors and flying qubits requires efficient and faithful QFC of single photons. In this letter we present, as a proof-of-principle experiment, quantum frequency down-conversion of visible single photons from an InP/GaInP QD using difference frequency generation (DFG) in a Zn-doped periodically poled Lithiumniobate (Zn:PPLN) waveguide (WG) \cite{Nishida2003}. The process is described by $1/\lambda_{\mathrm{in}} - 1/\lambda_{\mathrm{p}} = 1/\lambda_{\mathrm{out}}$, where $\lambda_{\mathrm{in}} = 711\,$nm is the wavelength of the input light field from a single InP/GaInP QD, $\lambda_{\mathrm{out}} = 1,313\,$nm is the wavelength of the down-converted output field, and $\lambda_{\mathrm{p}} = 1,550$~nm is the wavelength of a strong classical pump field. InP/GaInP QDs offer the opportunity of direct electrical excitation \cite{Reischle2010} which is advantageous for operation in optical communication systems. The long-wavelength pumping scheme is chosen to minimise noise background around $\lambda_{\mathrm{out}}$ due to Raman scattering \cite{Takesue2010, Curtz2011, Pelc2011, Zaske2011}.\\
\noindent Fig.~\ref{fig:Setup}a shows a schematic of our experimental setup which can be divided into three parts: a confocal microscope for investigation of the QD sample, a frequency conversion stage including the Zn:PPLN WG chip, and a spectral filtering stage for narrow spectral filtering of the downconverted signal. Using single photon input from a QD at a rate of $\sim \! 188,400$\,s$^{-1}$ we measure the efficiency $\eta_{\mathrm{QFC}}$ of our frequency conversion setup  as a function of the pump power $P_p$ at 1,550~nm that is coupled into the WG (Fig.~\ref{fig:Setup}b). We define $\eta_{\mathrm{QFC}} = N_{\mathrm{out}}/N_{\mathrm{in}}$, where $N_{\mathrm{in}}$ is the flux of visible photons at the entrance of the conversion stage and $N_{\mathrm{out}}$ is the flux of telecom photons at the output of WDM2 after the spectral filtering unit. For single photon detection at input and output side we use silicon avalanche photodiodes (Si-APDs) in the visible and NbN superconducting single-photon detectors (SSPDs) at telecom wavelengths, respectively (see methods for details). The data are fit using $\eta_{\mathrm{QFC}} (P_p) = \sin^2(\sqrt{\eta P_p} L)$, where $\eta=115$\,\%/(W\,cm$^2$) is a normalized efficiency and $L=4$\,cm is the length of the WG \cite{Roussev2004}. The point of maximum conversion is reached at $P_p = 150$~mW with $\eta^{\mathrm{(max)}}_{\mathrm{QFC}} \approx 0.32$, i.e., more than 30\% of the fraction of the quantum dot emission that can be coupled into a single-mode fibre is frequency down-converted in our setup. This corresponds to an internal efficiency for photons being converted inside the WG of $\eta_{\mathrm{int}} \gtrsim 0.64$. At the same time the ratio of converted photons to noise photons (signal-to-noise ratio, SNR) is larger than 20:1 (Fig.~\ref{fig:Setup}b). This unique combination of high conversion efficiency and high SNR enables faithful conversion of single photons. The acceptance bandwidth of the DFG process, as seen in Fig. \ref{fig:Setup}c, is determined to be $\Delta \lambda = 0.092$~nm or $\Delta \nu = 54.6$~GHz (full width at half maximum). Thus, the acceptance curve represents a narrow spectral bandpass filter inherent to the nonlinear process.\\
To illustrate the effect of spectral filtering Fig.~\ref{fig:FluorescenceMaps} shows a comparison of three PL maps along with their corresponding spectra. Fig.~\ref{fig:FluorescenceMaps}a and Fig.~\ref{fig:FluorescenceMaps}b were recorded without spectral filtering. Several QDs can be identified on the map of Fig.~\ref{fig:FluorescenceMaps}a. It is evident from the spectrum in Fig.~\ref{fig:FluorescenceMaps}b that a certain amount of background is still collected through the single-mode fibre. In order to select a single emission line from a single QD we insert a $56$~$\mu$m-thick silica etalon (Finesse $\mathcal{F} \approx 42$, free spectral range 1.85\,THz) as a narrow bandpass filter into the beamline of the visible PL light. This situation is represented in the map in Fig.~\ref{fig:FluorescenceMaps}c and the spectrum in Fig.~\ref{fig:FluorescenceMaps}d. The etalon clearly supresses PL from other QDs. The map in Fig.~\ref{fig:FluorescenceMaps}e was recorded by detecting converted IR-light employing a SSPD. In this case no etalon filtering of the visible input light was applied. However, from the IR-map and its related spectrum in Fig.~\ref{fig:FluorescenceMaps}f one recognizes that the acceptance bandwidth of the DFG process in combination with the FBG represents a narrow and efficient bandpass filter with high sideband suppression. No background or further emission lines are found in the IR spectrum.\\ 
Many implementations of quantum communication using single photons exploit their phase coherence \cite{Sangouard2011, Gisin2002}. In this context, it is crucial to conserve temporal coherence during the frequency conversion process. The degree of first-order coherence of a light field is given by $g^{(1)}(\tau) = \langle \hat{E}^-(t)\hat{E}^+(t + \tau) \rangle / \langle \hat{E}^-(t)\hat{E}^+(t) \rangle$, where $\hat{E}^-(t)$ and $\hat{E}^+(t)$ are electric-field operators and $\tau$ is a time delay. We measured $g^{(1)}(\tau)$ of the single photons and determined the coherence time $T_2$ employing a Michelson interferometer (see methods for details). The result is plotted in Fig.~\ref{fig:g1measurement}a for the 711~nm single photons emitted by the quantum dot. The visibility as a function of the time delay follows an exponential decay $\exp(-|\tau|/T_2)$ as expected for a Lorentzian lineshape of the QD emission. From the decay rate we find a coherence time of $T_2^{\mathrm{(vis)}} = 42 \pm 17$~ps. Interpreting the first-order coherence as an instantaneous linewidth, i.e. as short-term frequency fluctuations, we can give a worst case estimate of the coherence after frequency conversion: $\Delta \nu_{\mathrm{ir}} = \Delta \nu_{\mathrm{vis}} + \Delta \nu_{\mathrm{p}}$, where $\Delta \nu_{\mathrm{vis}}$, $\Delta \nu_{\mathrm{p}}$, and $\Delta \nu_{\mathrm{ir}}$ are the linewidth of the visible input field, the pump field and the converted field respectively. On short timescales the linewidth of the pump light $\Delta \nu_{\mathrm{p}}$ was determined to be far below 1~MHz corresponding to a long coherence time of $T_2^{\mathrm{(p)}} = 1/(2\pi \Delta \nu_{\mathrm{p}}) > 0.16\; \mu$s. As $T_2^{\mathrm{(p)}} \gg T_2^{\mathrm{(vis)}}$ the influence of the pump light on the coherence of the converted single photons is negligible. From the coherence measurement of the converted light (Fig.~\ref{fig:g1measurement}b) we find $T_2^{(\mathrm{IR})} = 49 \pm 13$~ps. Within the measurement error this value is in very good agreement with the coherence time of the original input photons proving conservation of the photon temporal coherence. We also measured the temporal width of the photons before and after conversion (Fig.~\ref{fig:g1measurement}c). The observed decay has two contributions determined by the lifetime of the excited state of the QD and a refilling process \cite{Aichele2004}. From the data we determine the lifetime of the QD emission to be $\tau_{\mathrm{vis}} = 2.6 \pm 0.1$~ns before and $\tau_{\mathrm{IR}} = 2.9 \pm 0.4$~ns after conversion. The results agree within the error bounds which is expected because the ultrafast conversion process should not add any temporal contribution. The much longer time constant of the refilling process in these measurements was about 2.5~$\mu$s.\\
We next investigate the preservation of the non-classical single photon statistics via the degree of second-order coherence $g^{(2)}(\tau) = \langle \hat{E}^-(t)\hat{E}^-(t + \tau) \hat{E}^+(t+\tau)\hat{E}^+(t) \rangle / \langle \hat{E}^-(t)\hat{E}^+(t) \rangle^2$ using Hanbury-Brown-Twiss (HBT) interferometers (see Methods). Fig. \ref{fig:g2measurement}a shows the $g^{(2)}$-function for the light emitted by the QD. The measured $g^{(2)}_{\mathrm{vis}}(0) = 0.39 < 0.5$ for the PL of the QD clearly proves single photon emission form a single quantum emitter. A Monte Carlo simulation of the emission process reproduces the measured data very well and reveals that $g^{(2)}_{\mathrm{vis}}(0) > 0$ due to the timing jitter of the APDs and uncorrelated photon emission from background passing the etalon filter ($\mathrm{SNR} \approx \mbox{7:1}$ determined from the spectrum of Fig.~\ref{fig:FluorescenceMaps}d). As represented in Fig. \ref{fig:g2measurement}b $g^{(2)}(\tau)$ was also measured for the down-converted light field. Here we obtain a value of $g^{(2)}_{\mathrm{IR}}(0) = 0.24$. The Monte Carlo simulation closely reproduces the measured data for $\mathrm{SNR}=\mbox{12:1}$ of the light field and a smaller timing jitter of the SSPD, indicating that the sub-Poissonian statistics of the input light field have been fully preserved as the SNR of the QFC process is much higher than the SNR of the light source. Even more, the SNR of the converted light is increased compared to the input light thanks to the strong spectral filtering effect of the QFC setup. The $g^{(2)}$ cross-correlation function (Fig.~\ref{fig:g2measurement}c) of original and converted photons exhibits a dip as well: $g^{(2)}_{\mathrm{vis/IR}}(\tau^{\prime}) = 0.44$. The observed anti-correlation again proves the conservation of the single particle character of the light upon QFC.\\
In summary, we have demonstrated efficient quantum frequency down-conversion of visible light emitted by a true single quantum emitter to a telecom wavelength. Strong spectral filtering enables low-noise operation of the converter and a high SNR. We have experimentally proven that the temporal coherence as well as photon antibunching, two essential prerequisites for applications in quantum communication and information, are conserved under frequency conversion. The converted single photons at 1,313\,nm (telecom O-band) are fibre coupled and the pump wavelength is at 1,550\,nm (telecom C-band) rendering our all-solid-state scheme fully compatible with existing telecom infrastructure. Combining a bright single quantum emitter in the visible with frequency down-conversion proves to be an elegant and flexible approach to realising a tunable telecom-band single photon source. Concerning the implementation of quantum repeaters we expect our experiment to work as well for input wavelengths relevant for alkali-atom-based quantum memories, e.g., 852\,nm (Cs D$_2$ line) \cite{Kuzmich2003} or 795/780\,nm (Rb D$_1$/D$_2$ line) \cite{Eisaman2005,Ritter2012}: chosing a pump wavelength between 1.9-2.1$\,\mu$m allows for low-noise down-conversion to one of the telecom spectral windows.\\

\pagebreak

\noindent \textbf{Methods:}\\
\noindent \textbf{Fabrication of InP/GaInP quantum dots.} The QD sample was grown by metal-organic vapor-phase epitaxy on a n-doped (100) GaAs substrate misorientated by 6$^\circ$ toward the (111)$_{A}$ direction. In order to enhance the collection efficiency an n-doped distributed Bragg reflector (DBR) of 10 $\lambda / 4$-pairs of Al$_{0.50}$Ga$_{0.50}$As/AlAs was placed below the active region. The InP QDs were grown self-assembled in the Stranski-Krastanow growth mode and were symmetrically embedded in 8.8\,nm intrinsic Ga$_{0.51}$In$_{0.49}$P barriers surrounded by 150\,nm partially doped (Al$_{0.55}$Ga$_{0.45}$)$_{0.51}$In$_{0.49}$P cladding layers. On top a p-doped aluminum rich oxidation layer was grown. The structure was terminated with a GaInP and a GaAs layer. After growth, standard processing techniques like lithography, thermal wet oxidation, and evaporation of ohmic contacts were applied to fabricate mesas of 100\,$\mu$m diameter. A ring shaped p-contact can be used to apply a DC bias voltage and the luminescence was collected inside a 20\,$\mu$m opening window of this ring contact. The QDs emit around 690-715\,nm and can be tuned by temperature and by Stark-effect via the bias voltage. For the experiment single QDs are optically addressed using the confocal microscope and a single QD with emission wavelength around 711\,nm has been selected.\\

\noindent \textbf{Pump source at 1,550~nm.} A home-built continuous wave optical parametric oscillator (similar to the one in \cite{Zaske2010}), pumped at 532~nm by a frequency doubled diode-pumped solid-state laser (Coherent Verdi V10), drives the DFG process. The OPO is based on a 30~mm-long periodically poled LiTaO$_3$ crystal in a four-mirror ring resonator which is resonant only for the signal wave. By variation of the quasi-phasematching (QPM) grating period and the temperature of the crystal the wavelength of the idler radiation can be set to any wavelength between 1,202~nm and 1,564~nm, corresponding to a signal wavelength of 806-954~nm. Depending on the output wavelength the OPO is able to deliver more than 1\,W of single-mode, single frequency output power with high intensity and frequency stability.\\

\noindent \textbf{Zn:PPLN waveguide chip.} The Zn:PPLN WG chip has dimensions of length $\times$ width $\times$ height = $40 \times 6 \times 0.5$~mm$^3$ and contains 12 ridge WGs with 6 different QPM grating periods $\Lambda_1 = 14.72, \Lambda_2 = 14.76,$ ... $, \Lambda_6 = 14.92$~$\mu$m. We choose $\Lambda_5 = 14.88$~$\mu$m and a temperature of $T_{\mathrm{WG}} = 28.05^{\circ}$C to achieve optimum QPM for the desired DFG process 1/710.74~nm $-$ 1/1,549.90~nm = 1/1,312.70~nm. The temperature of the crystal is controlled using a thermoelectric Peltier module.\\

\noindent \textbf{Michelson interferometer.} The Michelson interferometer consists of a 50/50 beamsplitter cube and retroreflecting prisms in each arm. The prisms' surfaces are uncoated to use them both in the visible as well as in the infrared spectral region. One of the retroreflectors is mounted on a long-range (30~cm) translation stage to set the time delay $\tau$ between the two arms. The other prism is mounted on a piezo actuator with 10~$\mu$m travel for the generation of interference fringes. The beamsplitter cube can be exchanged for the different spectral regions. The output of the interferometer was coupled into a single-mode fibre and guided to the appropriate detector (Si-APD for visible light, SSPD for telecom light). With the translation stage the path length difference is set to fixed values and at each position the piezo is scanned to observe interference fringes. The countrate of the detector is recorded during the piezo movement and is fitted with a sine-function. From this fit we can determine the minimum and maximum intensity in the fringes and calculate the visibility $V = (I_\mathrm{max}-I_\mathrm{min})/(I_\mathrm{max}+I_\mathrm{min})$.\\

\noindent \textbf{Photon counting and correlation measurements.} 
For single photon detection in the visible we use Si-APDs (Perkin Elmer SPCM-AQR-14) with 65\% detection efficiency around 700\,nm, dark count rate $\approx 300$\,s$^{-1}$, and 250\,ps timing jitter (standard deviation of Gaussian envelope). Infrared photons are detected by SSPDs (SCONTEL, Russia) consisting of an ultrathin (4\,nm thick, 100\,nm wide) NbN superconducting wire that is arranged to form a $10\,\mu$m $\times 10\,\mu$m meander structure. For photon detection, a bias current slightly below the critical current density is applied. The absorption of a photon leads then to the formation of a hot spot changing the current distribution in the wire. If the critical current density is exceeded the superconductivity is shortly lost over the width of the wire, leading to a voltage pulse that can be amplified and detected with a timing jitter below 25\,ps (standard deviation). A single mode fiber is mechanically fixed to the meander structure enabling the whole detector assembly to be put into a liquid helium storage dewar. By decreasing the pressure in the detector chamber the system can be operated at temperatures below 2\,K. Increasing the bias current raises both dark counts and quantum efficiency of the SSPDs. The optimal ratio has been found to be a setting with less than 10 dark counts per second. In this case, the quantum efficiency has been measured to be 12.2~$\pm$~0.7\,\% using an attenuated 1,310\,nm diode laser.\\
To determine the $g^{(2)}$-function for the light emitted directly by the quantum dot we use a Hanbury Brown-Twiss (HBT) interferometer consisting of a 50/50 beamsplitter cube and two Si-APDs. In the case of telecom light a fibre-based HBT with two SSPDs was used.
For all photon counting and correlation measurements we use a two channel time-correlated single photon counting (TCSPC) system (PicoHarp 300, PicoQuant, Germany). Each channel features a constant fraction discriminator (CFD). For lifetime measurements a TTL-signal synchronized with the optical pulses from the 590\,nm excitation source is input to channel 1. A single photon detector is connected to channel 2. Each TTL-pulse arriving at channel 1 starts a clock which is stopped by a detection event from the detector. The time intervals between start and stop events are recorded in a histogram. For the measurements of second-order photon correlation the TCSPC-modul is operated in time-tagged mode meaning that every detection event is stored in a list together with a time stamp. Using these lists software-assisted correlation of the events is performed on a PC. The output pulses of the Si-APDs associated with the detection of a photon are well adapted (concerning voltage level and shape) for the requirements of the TCSPC electronics. On the other hand, for the SSPD output pulses the CFD sometimes produces irregular pulse bursts. Due to channel cross-talk these pulse bursts appear on both input channels and thus can easily be identified and discarded by the data acquisition software.

\noindent \textbf{Acknowledgements}\\
We are indebted to Carsten Kremb for his quick help with wire bonding. We further thank Felix R\"ubel and Johannes L'huillier (J.L.) for lending us the fs-OPO and J.L. for helpful discussions in the early stage of the project. This work was financially supported by the Bundesministerium f\"ur Bildung und Forschung (joint project between networks QuOReP, Contract~No.~01BQ1011, QuaHL-Rep, Contract~No.~01BQ1041, and EphQuaM, Contract~No.~01BL0902).\\

\noindent \textbf{Author Contributions}\\
S.Z. constructed the frequency conversion setup. A.L. built the Michelson interferometer. C.A.K. performed preliminary investigations of the QD sample. J.K. operated the SSPDs. C.H. and C.A. operated the confocal microscopy setup. R.A. designed and built the cryogenic part of the confocal microscope and helped with data acquisition. W.-M.S. and M.J. designed the QD sample structure and carried out epitaxial growth. S.Z., A.L., C.H., C.A., C.A.K. and J.K. conducted the experiments. S.Z. and A.L. analysed the data. P.M. and C.B. planned and supervised the project. S.Z. and C.B. wrote the manuscript. All authors discussed the results and commented on the manuscript.\\

\noindent \textbf{Competing Financial Interests}\\
The authors declare no competing financial interests.

\end{document}